\documentclass[aps,prl,twocolumn,showpacs,superscriptaddress,floatfix]{revtex4-1}

\usepackage{amsmath,amssymb}
\usepackage{graphicx}
\usepackage{epsfig}
\usepackage[none]{hyphenat}
\begin{document}

\title{Suppression of Superconductivity by Twin Boundaries in FeSe}

\author{Can-Li Song}
\affiliation{State Key Laboratory for Surface Physics, Institute of Physics, Chinese Academy of Sciences, Beijing 100190, China}\affiliation{State Key Laboratory of Low-Dimensional Quantum Physics, Department of Physics, Tsinghua University, Beijing 100084, China}\affiliation{Department of Physics, Harvard University, Cambridge, MA 02138, U. S. A}
\author{Yi-Lin Wang}
\affiliation{State Key Laboratory for Surface Physics, Institute of Physics, Chinese Academy of Sciences, Beijing 100190, China}
\author{Ye-Ping Jiang}
\affiliation{State Key Laboratory for Surface Physics, Institute of Physics, Chinese Academy of Sciences, Beijing 100190, China}\affiliation{State Key Laboratory of Low-Dimensional Quantum Physics, Department of Physics, Tsinghua University, Beijing 100084, China}
\author{Lili Wang}
\author{Ke He}\affiliation{State Key Laboratory for Surface Physics, Institute of Physics, Chinese Academy of Sciences, Beijing 100190, China}
\author{Xi Chen}
\affiliation{State Key Laboratory of Low-Dimensional Quantum Physics, Department of Physics, Tsinghua University, Beijing 100084, China}
\author{Jennifer E. Hoffman}
\affiliation{Department of Physics, Harvard University, Cambridge, MA 02138, U. S. A}
\author{Xu-Cun Ma}
\email[]{xcma@aphy.iphy.ac.cn}
\affiliation{State Key Laboratory for Surface Physics, Institute of Physics, Chinese Academy of Sciences, Beijing 100190, China}
\author{Qi-Kun Xue}
\email[]{qkxue@mail.tsinghua.edu.cn}
\affiliation{State Key Laboratory for Surface Physics, Institute of Physics, Chinese Academy of Sciences, Beijing 100190, China}\affiliation{State Key Laboratory of Low-Dimensional Quantum Physics, Department of Physics, Tsinghua University, Beijing 100084, China}
\date{\today}

\begin{abstract}
Low-temperature scanning tunneling microscopy and spectroscopy are employed to investigate twin boundaries in stoichiometric FeSe films grown by molecular beam epitaxy. Twin boundaries can be unambiguously identified by imaging the 90$^{\circ}$ change in the orientation of local electronic dimers from Fe site impurities on either side. Twin boundaries run at approximately 45$^{\circ}$ to the Fe-Fe bond directions, and noticeably suppress the superconducting gap, in contrast with the recent experimental and theoretical findings in other iron pnictides. Furthermore, vortices appear to accumulate on twin boundaries, consistent with the degraded superconductivity there. The variation in superconductivity is likely caused by the increased Se height in the vicinity of twin boundaries, providing the first local evidence for the importance of this height to the mechanism of superconductivity.
\end{abstract}


\maketitle

The response of superconductivity to crystal defects is crucial to two forefront technological issues, namely the sharpness of the superconducting transition and the critical current. Many early studies have revealed a slight enhancement in the superconducting critical temperature $\textit{T}_c$ near twin boundaries (TBs) of certain conventional superconductors such as In, Sn, and Nb \cite{khlyustikov1987twinning}. Meanwhile, TBs tend to pin vortices and so enhance the critical currents in the cuprate high-$\textit{T}_c$ superconductor $\textrm{YBa}_{2}\textrm{Cu}_{3}\textrm{O}_{7-\delta}$ (YBCO) \cite{dolan1989vortex,maggio1997critical}. The general interplay of TBs and superconducting properties remains unresolved.

In the recently discovered iron-based compounds, the tetragonal-to-orthorhombic distortion above $\textit{T}_c$ typically generates a maze of TBs upon cooling \cite{tanatar2009direct}, which serves as a test bed for twinning-plane superconductivity. Local susceptometry measurements with a scanned superconducting quantum interference device (SQUID) show an enhanced superfluid density along TBs in underdoped $\textrm{Ba}(\textrm{Fe}_{1-x}\textrm{Co}_{x})_{2}\textrm{As}_{2}$ \cite{kalisky2010stripes,kirtley2010meissner}, compatible with SQUID magnetometry images where vortices avoid pinning on TBs \cite{kalisky2011behavior}. On the other hand, doping-dependent TB imaging with polarized light, combined with bulk critical current determination in $\textrm{Ba}(\textrm{Fe}_{1-x}\textrm{Co}_{x})_{2}\textrm{As}_{2}$, shows a tremendous enhancement of critical current at the doping level where TBs are densest, leading to a claim that vortices are pinned on TBs \cite{prozorov2009intrinsic}. However, in the absence of direct vortex imaging, this latter observation is also consistent with the possibility that the critical current is enhanced by vortex trapping between the TBs. Bitter decoration in $\textrm{Ba}(\textrm{Fe}_{1-x}\textrm{Ni}_{x})_{2}\textrm{As}_{2}$ shows vortices clustered along lines in some regions of the sample, leading to a claim of TB pinning \cite{li2011low}. But the absence of simultaneous twin boundary imaging again leaves open the possibility that the aligned vortices are pinned on  domains between parallel TBs.

These experiments on iron-based superconductors have used magnetic imaging techniques, whose resolution is limited to approximately the penetration depth $\lambda\sim325$ nm \cite{luan2010local}. Because pinning may occur on the vortex core length scale, $\xi\sim3$ nm \cite{yin2009scanning}, it can be challenging in some cases to determine from magnetic imaging alone whether a vortex is pinned on or near the TB. Scanning tunneling microscopy (STM) and spectroscopy (STS), which can image both TBs and vortices on the $\xi$  length scale, can address this issue with a resolution $\times100$ better \cite{hoffman2011spectroscopic}. Additionally, in most cuprates and iron pnictides, chemical doping plays an essential role in superconductivity; thus its possible variation across TBs may complicate the understanding of the twinning-plane superconductivity \cite{yan1996changes}. As an alternative, the stoichiometric and structurally simple PbO-type $\beta$-FeSe superconductor provides a unique system for addressing the variations in superconductivity near TBs \cite{hsu2008superconductivity}.

Here we report on STM and STS studies of TBs in stoichiometric and superconducting FeSe films grown by molecular beam epitaxy (MBE). This allows for a direct probe of the superconducting order parameter near TBs at the nanometer length scale. Sparse Se atoms at near-surface Fe sites produce local dimerlike scattering signatures. TBs are identified by the $90^{\circ}$ rotations of the electronic dimers on either side, and are seen to roughly orient along the diagonals of the Fe unit cells. We observe that (i) TBs considerably suppress the superconducting gap within the coherence length, and (ii) vortices tend to be pinned on TBs. Both observations demonstrate that TBs locally weaken the superconductivity in FeSe.

All STM and STS tunneling experiments presented here were carried out at 4.5 K on a commercial ultrahigh vacuum low temperature STM apparatus (Unisoku), which is connected to a MBE system for \textit{in situ} sample preparation. The base pressure for both systems is better than $10^{-10}$ Torr. The MBE growth of stoichiometric FeSe films has been described in detail elsewhere \cite{song2011direct,song2011molecular}, and in the supplemental Material \cite{supplementary}. Prior to data collection, a polycrystalline PtIr tip was cleaned by electron-beam heating in ultrahigh vacuum, and then calibrated on a MBE-grown Ag film on a Si(111) substrate. Spectroscopic measurements were made by disrupting the feedback circuit, sweeping the sample voltage, and extracting the tunneling conductance \textit{dI/dV} using a standard lock-in technique with a small bias modulation of 0.1 mV at 987.5 Hz.

\begin{figure}[tbh]
\includegraphics[width=\columnwidth]{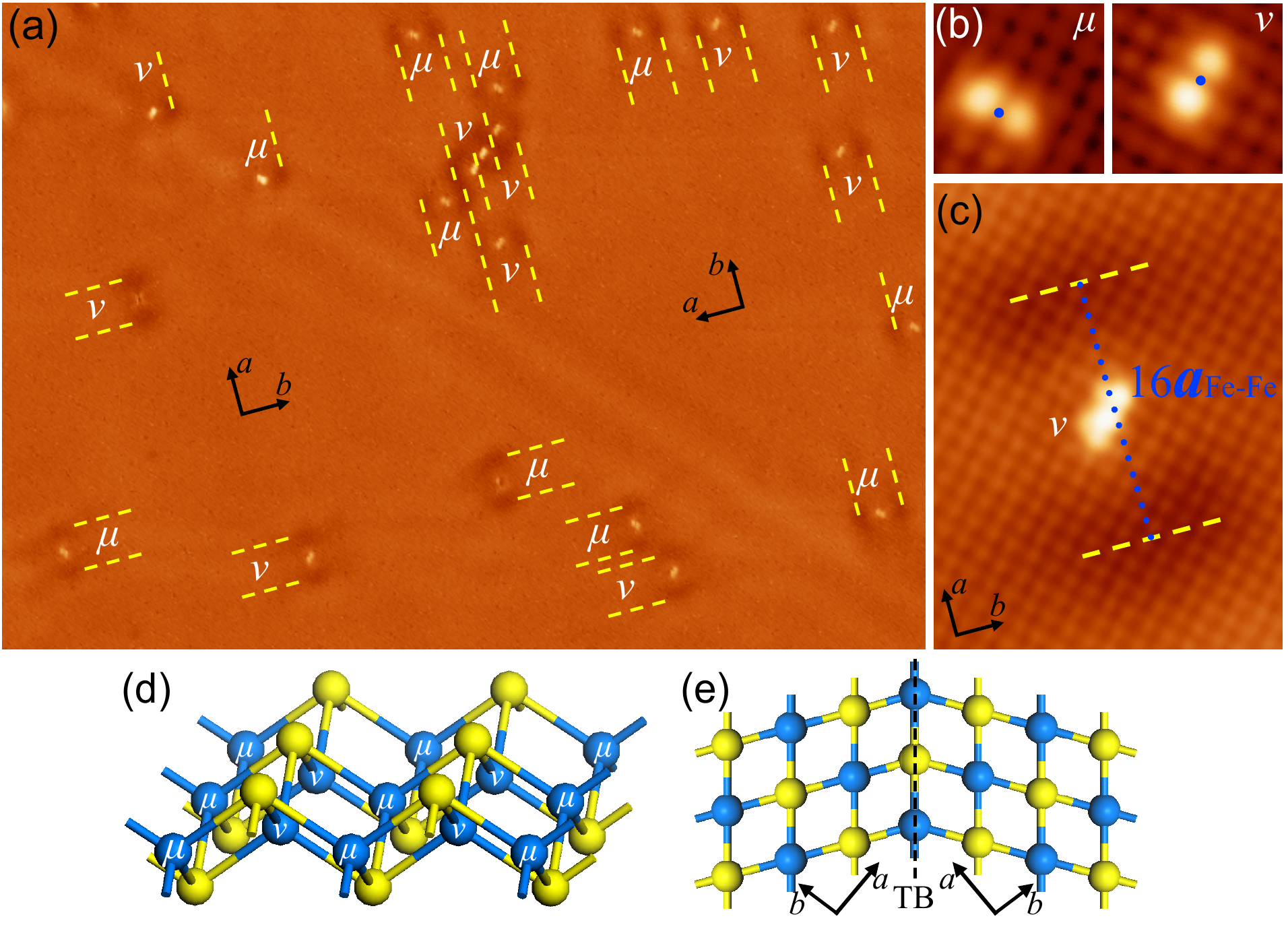}
\caption{(color online) (a) STM topography of FeSe film with extra Se atoms appearing as bright atomic-scale dumbbells (\textit{V} = 10 mV, \textit{I} = 0.1 nA, 100 nm $\times$ 70 nm). (b) Zoom-in on two orthogonally oriented atomic dumbbells, labeled $\mu$ and $\nu$ (\textit{V} = 6 mV, \textit{I} = 0.1 nA, 2 nm $\times$ 2 nm). (c) Larger zoom-in of a single excess Se atom (\textit{V} = 10 mV, \textit{I} = 0.1 nA, 6 nm $\times$ 8 nm). The blue dots mark the subsurface Fe atoms. The depressions straddling each excess Se (marked by dashed yellow lines) and the TB likely stem from quasiparticle scattering. (d) Schematic crystal structure of $\beta\textrm{-FeSe}$ showing the inequivalent $\mu$ and $\nu$ Fe sites, and (e) diagram illustrating a TB with Fe (blue) and Se (yellow) spheres.}
\end{figure}

Figure 1(a) depicts a constant-current topographic image of an as-grown FeSe films. The localized defects ($<0.05\%$) correspond to individual excess Se atoms \cite{song2011molecular}, which are intentionally introduced and act as scatterers for electrons and give rise to unidirectional electronic nanostructures in FeSe \cite{song2011direct}. A more detailed examination shows that each excess Se explicitly breaks fourfold ($C_{4}$) rotational symmetry in two independent ways, at two different length scales. First, at the atomic length scale, we observe two orthogonal dumbbell-like features, labeled as $\mu$ and $\nu$. Both atomic dumbbells are centered at subsurface Fe atoms [Fig.~1(b)],
with their bright ends positioned on two adjacent Se atoms in the topmost layer, suggesting that the excess Se substitutes into the uppermost Fe layer. Two inequivalent Fe positions, denoted by $\mu$ and $\nu$ in Fig.~1(d), lead to the two orthogonal atomic dumbbells observed. Second, at the much larger length scale of $\sim16a_{\textrm{Fe-Fe}} \backsimeq 4.4 $ nm [Fig.~1(c)], the $C_{4}$ symmetry is broken by unidirectional depressions in the density of states which straddle each excess Se (yellow dashes). In contrast to the persistence of atomic dumbbells up to 2.5 eV imaging bias, the larger unidirectional features exist only in a narrow energy range (approximately $\pm20$ meV), which supports a purely electronic origin. Note that in Fig.~1(a) a faint stripe occurs along the upper left to lower right diagonal. Across this stripe, the Se-induced unidirectional nanostructures (electronic dimers) are found to rotate by $90^{\circ}$. This closely resembles the TB-induced 90$^{\circ}$ rotation of $\sim8a_{\textrm{Fe-Fe}}$ unidirectional nanostructures in slightly Co-doped $\textrm{Ca}(\textrm{Fe}_{1-x}\textrm{Co}_{x})_{2}\textrm{As}_{2}$ \cite{chuang2010nematic}. We therefore argue that the observed faint stripe along the diagonal of Fig.~1(a) represents a TB, across which the \textit{a} and \textit{b} crystalline axes interchange. Here \textit{a} and \textit{b} correspond to the two Fe-Fe bonding directions, as defined in Fig.~1(e). Note that these larger electronic dimers always respect the crystalline \textit{a} axis, irrespective of the stochastic distribution of atomic {$\mu$} and {$\nu$} dumbbells. These observations not only provide a way to distinguish TBs, but also support the fundamental role that electronic dimers play in scattering mechanisms \cite{Allan2012Dimer,Kang2012dimer}, and by implication, transport anisotropy in iron pnictides \cite{chu2010plane}.

\begin{figure}[tbh]
 \includegraphics[width=0.99\columnwidth]{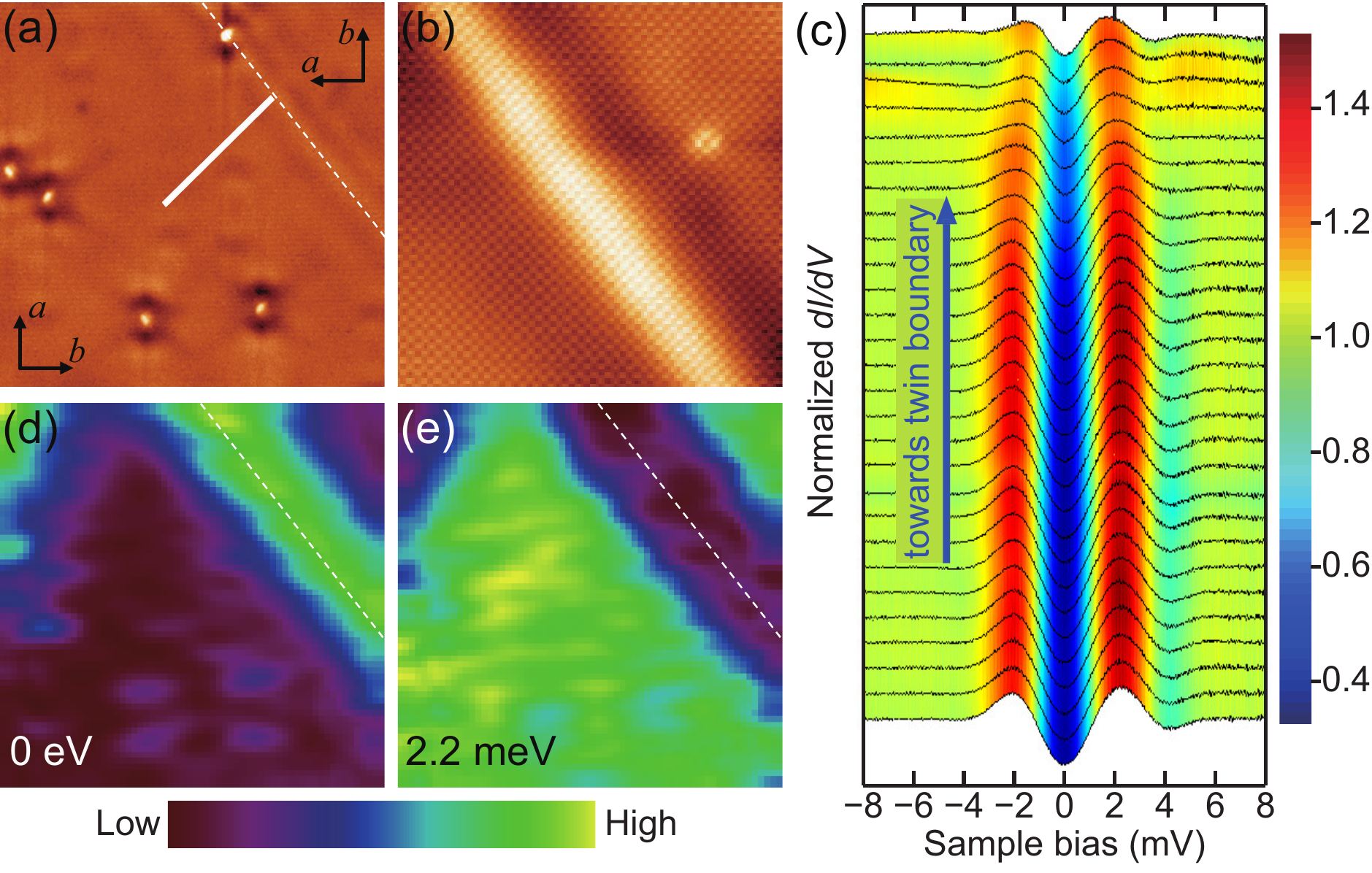}
  \caption{(color online) (a) STM topography with a TB indicated by a white dashed line (\textit{V} = 10 mV, \textit{I} = 0.1 nA, 50 nm $\times$ 50 nm). (b) Atomically resolved topography of a TB (\textit{V} = 10 mV, \textit{I} = 43 pA, 12 nm $\times$ 12 nm). (c) Normalized \textit{dI/dV} spectra taken at equal separations (1 nm) along the white solid line (normal to the TB) in (a). (d, e) Differential conductance maps recorded simultaneously with image (a) at energies of (d) zero and (e) 2.2 meV, respectively. Tunneling gap is set at \textit{V} = 10 mV and \textit{I} = 0.1 nA.}
\end{figure}

To find the effect of TBs on superconductivity, we use STS to map the superconducting gap in the vicinity of another TB [Fig.~2(a)]. Using the atomically-resolved STM image in Fig.~2(b), we note that the TB runs nearly along one of the Se-Se nearest-neighbor directions in the topmost layer, or equivalently one diagonal of the undistorted single-Fe unit cells. Figure 2(c) shows a series of differential conductance \textit{dI/dV} spectra, normalized to the normal-state conductance spectrum above $\textit{T}_c$ (10 K), taken along a trajectory approaching the twin boundary. All curves exhibit superconducting gaps with clear coherence peaks. However the gap magnitude $\Delta$, half of the energy between the coherence peaks, decreases when approaching the TB, suggesting that TBs tend to weaken the superconductivity in FeSe. This is further supported by \textit{dI/dV} maps at zero energy [Fig.~2(d)] and at one of the coherence peaks at $\sim2.2$ meV [Fig.~2(e)] on the same region as Fig.~2(a). The TB enhances the zero-bias conductance (ZBC, inversely correlated with the superfluid density) and suppresses the coherence peaks. Our observations consistently support the suppression of superconductivity by TBs in FeSe. This contrasts with the enhanced superfluid density along TBs in underdoped $\textrm{Ba}(\textrm{Fe}_{1-x}\textrm{Co}_{x})_{2}\textrm{As}_{2}$ by SQUID measurements as well as the recent theoretical prediction \cite{kalisky2010stripes,huang2011domain}.

\begin{figure}[tbh]
  \includegraphics[width=.8\columnwidth]{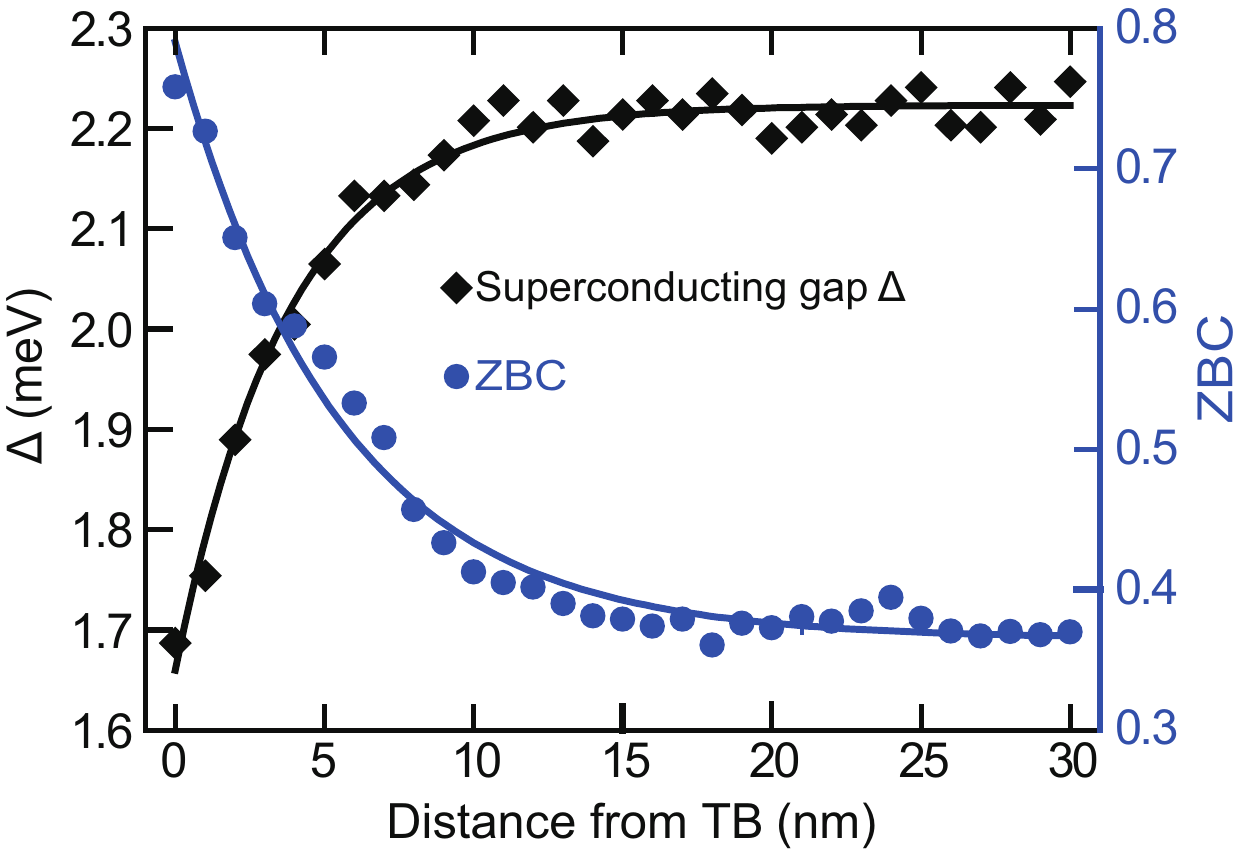}
  \caption{(color online) Superconducting gap $\Delta$ (black diamonds) and ZBC (blue circles) plotted as a function of distance from the TB. The blue solid line depicts an exponential decay, while the black one is a guide to the eye.}
\end{figure}

Figure 3 presents the extracted superconducting gap $\Delta$ and ZBC from Fig.~2(c) as a function of distance \textit{d} off the TB. As compared to $\Delta_{0}$ = 2.2 meV on TB-free regions, the superconducting gap shrinks by $\sim25\%$ to $\Delta_{\textrm{TB}}$ = 1.66 meV on TBs. Also, ZBC(\textit{d}) decays with distance \textit{d} from the TB as $\textrm{ZBC}(\textit{d}) = \textrm{ZBC}(\infty ) + A\textrm{exp}(-\textit{d}/\xi$). Here $\textrm{ZBC}(\infty)$ and $\xi$  are the constant background and superconducting coherence length, respectively. Based on the exponential fitting, we extract a coherence length of $\xi = 5.5\pm0.3$ nm at 4.5 K. The coherence length $\xi{(0)}\sim 5.1$ nm at zero temperature can be calculated from the self-consistent BCS gap function and $\xi{(T)}\propto1/\Delta{(T)}$ \cite{tinkham2004introduction} with $\textit{T}_c \simeq 9.3$ K \cite{song2011molecular}. We note that anisotropic vortices have been recently demonstrated in FeSe and can be intuitively understood by direction-dependent changes in $\xi$ with extrema along the $a$ and $b$ directions \cite{song2011direct}. In this work, $\xi$ is measured along one diagonal of the undistorted Fe unit cells, and thus $\xi \sim 5.1\,\mathrm{nm}$ roughly represents an average of $\xi_a$ and $\xi_b$, which is comparable to the estimated coherence length of 4.5 nm from transport measurements \cite{hsu2008superconductivity}. Our STM images therefore demonstrate directly for the first time the coherence-length-scale effect of a TB on superconductivity in the new Fe-based superconductors.

\begin{figure}[tbh]
 \includegraphics[width=\columnwidth]{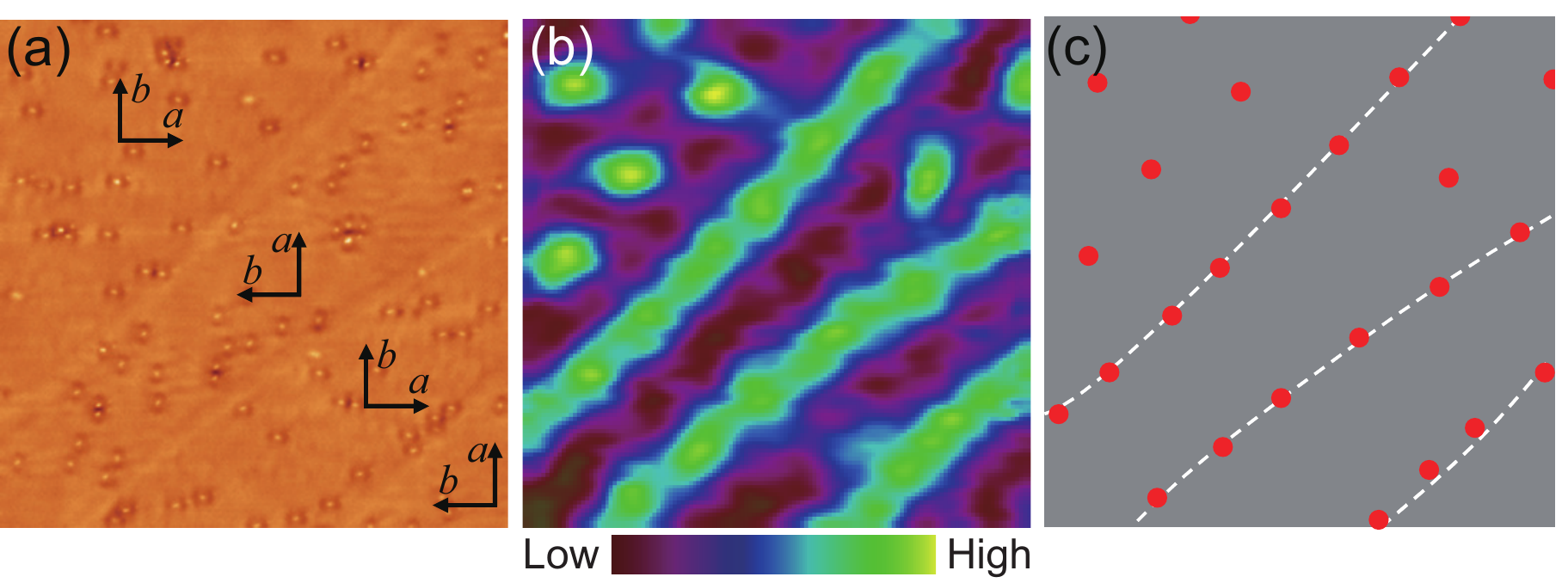}
  \caption{(color online) (a) A 150 nm $\times$ 150 nm topographic image with three TBs (\textit{V} = 10 mV, \textit{I} = 0.1 nA). (b) Simultaneous ZBC map showing the vortices at 2 T. Tunneling gap is set at \textit{V} = 10 mV and \textit{I} = 0.1 nA. (c) Schematic illustrating both TBs (white dashes) and vortices (red circles).}
\end{figure}

The suppressed superconductivity and thus reduced superfluid density along TBs should lead to a decrease in energy when vortices are positioned on TBs. To search for such TB flux pinning, we image vortices with an applied magnetic field normal to the FeSe \textit{ab}-plane. Figure 4(a) shows a topographic image with three TBs, where we record the ZBC map at 2 T, illustrated in Fig.~4(b). Previously, a pronounced ZBC peak, which originates from quasiparticle bound states \cite{caroli1964bound,shore1989density}, has been found in the vortex cores of FeSe \cite{song2011direct}. Therefore the yellow regions with enhanced ZBC signify individual isolated vortices. The observed average flux per vortex is $\sim2.05 \times 10^{-15}$ Wb \cite{supplementary}, consistent with single magnetic flux quantum, $\Phi_{0} = 2.07 \times 10^{-15}$ Wb. The schematic depiction of vortices and twin boundaries in Fig.~4(c) illustrates that vortices are preferentially pinned on TBs \cite{supplementary} as long as the distance separating the neighboring TBs is not too large. This observation confirms that TBs locally suppress the superconductivity in FeSe.

Now we consider possible explanations for the suppressed superconductivity by TBs in FeSe. A variation in chemical doping across TBs can be reasonably excluded \cite{yan1996changes} because the superconductivity develops in FeSe without any external doping, in sharp contrast to iron pnictides \cite{johnston2010puzzle}. We thus consider that the phenomenon likely stems from the structural changes around TBs. Indeed, in iron-based superconductors, the tetrahedral geometry, both the tetrahedral angle $\alpha$, and the anion height $h_{\textrm{anion}}$(pnictogen or chalcogen) above the Fe layer, appear to be key parameters controlling the superconducting transition temperature $\textit{T}_c$ \cite{johnston2010puzzle,huang2010control,horigane2009relationship,okabe2010pressure,mizuguchi2010anion}. For each $\textrm{FeSe}_{4}$ tetrahedron spanning a TB, two out of four Se anions must be mirror symmetric with respect to the twinning plane [Fig.~1(e)], which will distort the $\textrm{FeSe}_{4}$ tetrahedra and thus change $\alpha$. However, some previous studies have demonstrated that $\alpha$ does not significantly affect $\textit{T}_c$ in iron chalocogenides \cite{huang2010control,horigane2009relationship}. We therefore suggest that the tetrahedral distortion cannot bear sole responsibility for the observed suppression of superconductivity around TBs in FeSe. Then we examine the Se height $h_{\textrm{Se}}$ around TBs. High-pressure electrical resistivity measurements revealed an enhanced $\textit{T}_c$  as $h_{\textrm{Se}}$ is reduced \cite{okabe2010pressure}. In all our topographic images, up to 1 eV \cite{supplementary}, TBs appear brighter than surrounding areas. This consistency over a wide energy range strongly suggests a local increase in $h_\mathrm{Se}$, although an electronic effect leading to the false appearance of increased height due to the STM normalization artifact cannot be completely ruled out. We therefore suggest that superconductivity is suppressed and perhaps even quenched by the increased $h_{\mathrm{Se}}$ at the TB. The well-identified superconducting gaps near TBs [Fig.~2(c)] may arise from the proximity effect between on- and off-TBs regions.

Finally we tentatively explain the contrasting roles of TBs in $\textrm{Ba}(\textrm{Fe}_{1-x}\textrm{Co}_{x})_{2}\textrm{As}_{2}$ and FeSe. As has been noted previously, the $\textit{T}_c$ of the iron-based superconductors appears to reach a maximum at $h_{\textrm{anion}} \backsimeq 1.38$ ${\textrm{\AA}}$ \cite{johnston2010puzzle,okabe2010pressure,mizuguchi2010anion}. Away from this value, $\textit{T}_c$ will abruptly decrease. In FeSe, $h_{\textrm{Se}} \simeq 1.45$ ${\textrm{\AA}} > 1.38$ ${\textrm{\AA}}$ \cite{okabe2010pressure}, so the increased $h_{\textrm{Se}}$ must suppress $\textit{T}_c$ at TBs. However, in $\textrm{Ba}(\textrm{Fe}_{1-x}\textrm{Co}_{x})_{2}\textrm{As}_{2}$, $h_{\textrm{As}} \backsimeq 1.34$ ${\textrm{\AA}}$ appears smaller than 1.38 ${\textrm{\AA}}$ \cite{drotziger2010pressure}. Assuming that $h_{\textrm{As}}$ increases around TBs of $\textrm{Ba}(\textrm{Fe}_{1-x}\textrm{Co}_{x})_{2}\textrm{As}_{2}$ as well, one can expect enhanced superfluid density there, in line with SQUID experiments \cite{kalisky2010stripes,kirtley2010meissner}. Moreover, $h_{\textrm{anion}}$ may play a more important role in FeSe than in $\textrm{Ba}(\textrm{Fe}_{1-x}\textrm{Co}_{x})_{2}\textrm{As}_{2}$. The increased $h_{\textrm{Se}}$ at FeSe TBs favors the double-stripe ($\pi$, 0) magnetic order, and suppresses the ($\pi$, $\pi$) spin fluctuations which are necessary for superconductivity \cite{moon2010chalcogen}. The present study therefore provides evidence linking $h_{\textrm{anion}}$ to the local $\Delta$ and thus to the mechanism of superconductivity in iron-based compounds.

Our detailed STM and STS study of TBs in MBE-grown FeSe films has provided fundamental new information about the nature of superconductivity in iron-based materials. First, we have explicitly shown by direct imaging that each Fe-site impurity produces a local electronic dimer of size $\sim16a_{\textrm{Fe-Fe}}$, oriented along the orthorhombic \textit{a} axis. Scattering from these dimers, although never previously directly visualized in real space, has been controversially suggested as the root cause of the transport anisotropy in iron-based superconductors \cite {Allan2012Dimer,Kang2012dimer}. Second, we have shown by spatially resolved spectroscopy that TBs suppress the superconductivity within a superconducting coherence length $\xi$. This provides a quantitative measure of the coherence length,  $\xi\sim 5.1$ nm. Third, we show that magnetic vortices are preferentially pinned to the TBs. This supports the suppression of superconductivity at the TBs, and can inform engineering work to optimize vortex pinning for increased critical current. Finally, we show increased $h_{\textrm{Se}}$ at the FeSe TBs. This suggests an explanation for the contrast between TB behavior in FeSe vs. $\textrm{Ba}(\textrm{Fe}_{1-x}\textrm{Co}_{x})_{2}\textrm{As}_{2}$, and indeed provides the first local evidence for the importance of chalcogen or pnictogen height $h_{\mathrm{anion}}$ to the very nature of the superconducting mechanism in iron-based materials.

\begin{acknowledgments}
 This work was supported by National Science Foundation and Ministry of Science and Technology of China. C. L. S was supported by the Golub Fellowship at Harvard University. Topographic images were partly processed with the WSxM software \cite{horcas2007wsxm}.
\end{acknowledgments}

%

\end{document}


\Large
\begin{spacing}{1}
{\textbf{Supplemental Material for:}}
\end{spacing}
\title{Suppression of Superconductivity by Twin Boundaries in FeSe}
\small
\author{Can-Li Song}
\affiliation{State Key Laboratory for Surface Physics, Institute of Physics, Chinese Academy of Sciences, Beijing 100190, China}\affiliation{State Key Laboratory of Low-Dimensional Quantum Physics, Department of Physics, Tsinghua University, Beijing 100084, China}\affiliation{Department of Physics, Harvard University, Cambridge, MA 02138, U. S. A}
\author{Yi-Lin Wang}
\affiliation{State Key Laboratory for Surface Physics, Institute of Physics, Chinese Academy of Sciences, Beijing 100190, China}
\author{Ye-Ping Jiang}
\affiliation{State Key Laboratory for Surface Physics, Institute of Physics, Chinese Academy of Sciences, Beijing 100190, China}\affiliation{State Key Laboratory of Low-Dimensional Quantum Physics, Department of Physics, Tsinghua University, Beijing 100084, China}
\author{Lili Wang}
\author{Ke He}\affiliation{State Key Laboratory for Surface Physics, Institute of Physics, Chinese Academy of Sciences, Beijing 100190, China}
\author{Xi Chen}
\affiliation{State Key Laboratory of Low-Dimensional Quantum Physics, Department of Physics, Tsinghua University, Beijing 100084, China}
\author{Jennifer E. Hoffman}
\affiliation{Department of Physics, Harvard University, Cambridge, MA 02138, U. S. A}
\author{Xu-Cun Ma}
\email[]{xcma@aphy.iphy.ac.cn}
\affiliation{State Key Laboratory for Surface Physics, Institute of Physics, Chinese Academy of Sciences, Beijing 100190, China}
\author{Qi-Kun Xue}
\email[]{qkxue@mail.tsinghua.edu.cn}
\affiliation{State Key Laboratory for Surface Physics, Institute of Physics, Chinese Academy of Sciences, Beijing 100190, China}\affiliation{State Key Laboratory of Low-Dimensional Quantum Physics, Department of Physics, Tsinghua University, Beijing 100084, China}

\begin{abstract}
\end{abstract}

\maketitle

We prepare high-quality FeSe films on a graphitized SiC(0001) substrate \cite{song2011molecular}. The substrate, with a resistivity of $\sim 0.1\,\Omega\cdot\mathrm{cm}$, consists mainly of double-layer graphene, which is prepared using a well-established recipe \cite{hass2008growth}. We co-evaporate high purity Fe ($99.995\%$) and Se ($99.999\%$) sources from standard Knudsen cells onto the substrate at $450^{\circ}$C. During the film growth, the nominal Se/Fe beam flux ratio is $\textrm{20}$, and the pressure is $10^{-9}$ Torr, due to the volatile Se molecular beam. The high Se/Fe flux ratio compensates the losses of volatile Se molecules, and leads to stoichiometric FeSe films. The excess Se will desorb and cannot be incorporated into the stoichiometric FeSe due to the high substrate temperature.

The Fe$_{1-x}$Se$_{1+x}$ samples (where $x$ refers to the doping level in the top Se-Fe$_2$-Se layer) were prepared by depositing additional Se onto the stoichiometric FeSe films held at the lower substrate temperature of $220^{\circ}$C, where the excess Se were found to incorporate into the topmost Fe sites. Precise control of Se dosage enabled us to grow Fe$_{1-x}$Se$_{1+x}$ films at various $x$ up to $x$ = $10\%$. We found that the superconductivity was completely destroyed at a doping level of approximately $x$ = $2.5\%$. In the present study, the small $x<0.05\%$ barely disturbs the superconductivity. For example, Fig.\ 1(a) contains 21 excess Se atoms in a 70 nm $\times$ 100 nm field of view, corresponding to $x=0.02\%$; Fig.\ 2(a) contains 5 excess Se atoms in a 50 nm $\times$ 50 nm field of view, corresponding to $x=0.014\%$; Fig.\ 4(a) contains 99 excess Se atoms in a 150 nm $\times$ 150 nm field of view, corresponding to $x=0.032\%$.

Figure 1 shows the vortex density in various magnetic fields (1, 2, 4 and 8 Tesla). The density depends linearly on the field and agrees well with the theoretical expectation.
\begin{figure}[tbh]
\includegraphics[width=0.52\columnwidth]{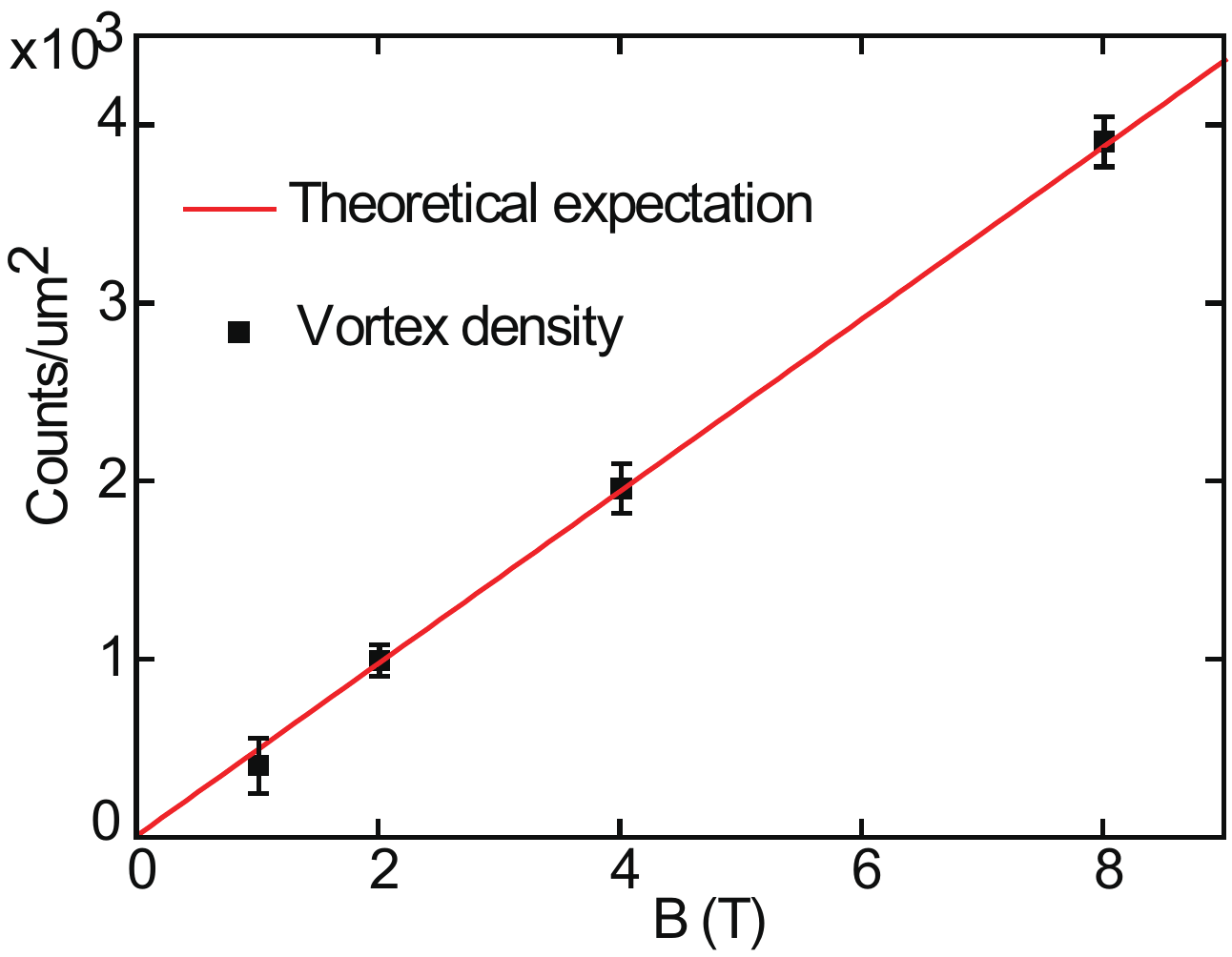}
\caption{ Magnetic field dependence of vortex density in FeSe.}
\end{figure}

To justify the exact position of vortex cores, we simultaneously imaged the high-resolution vortices and STM topography, as shown in Fig.\ 2. Here, it is clear that the vortex cores are located in the bright central region, rather than the two dark trenches.
\begin{figure}[tbh]
\includegraphics[width=1\columnwidth]{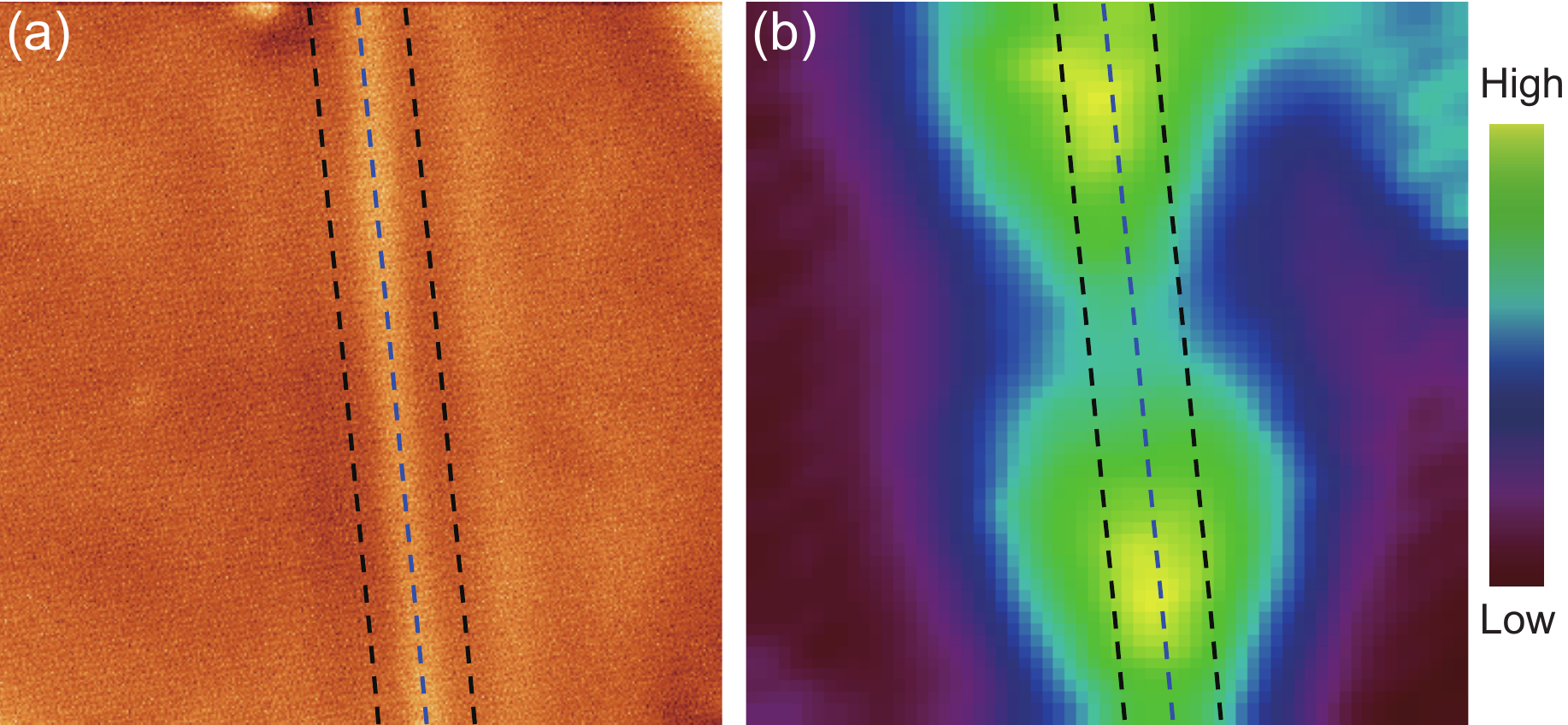}
  \caption{ (a) STM topography of a TB (\textit{V} = 10 mV, \textit{I} = 0.1 nA, 40 nm $\times$ 40 nm), and (b) the simultaneously acquired ZBC map showing the vortices at 2 T. The tunneling gap is set at \textit{V} = 10 mV and \textit{I} = 0.1 nA. The blue and black dashes indicate the TB and the two dark trenches straddling the TB, respectively.}
\end{figure}

Figure 3 shows the STM topography of a TB at high voltage (1 V). Even at high voltage, the TB
appears brighter than the surrounding plane, indicating a true geometric height
elevation rather than a purely electronic effect.
\begin{figure}[tbh]
\includegraphics[width=1\columnwidth]{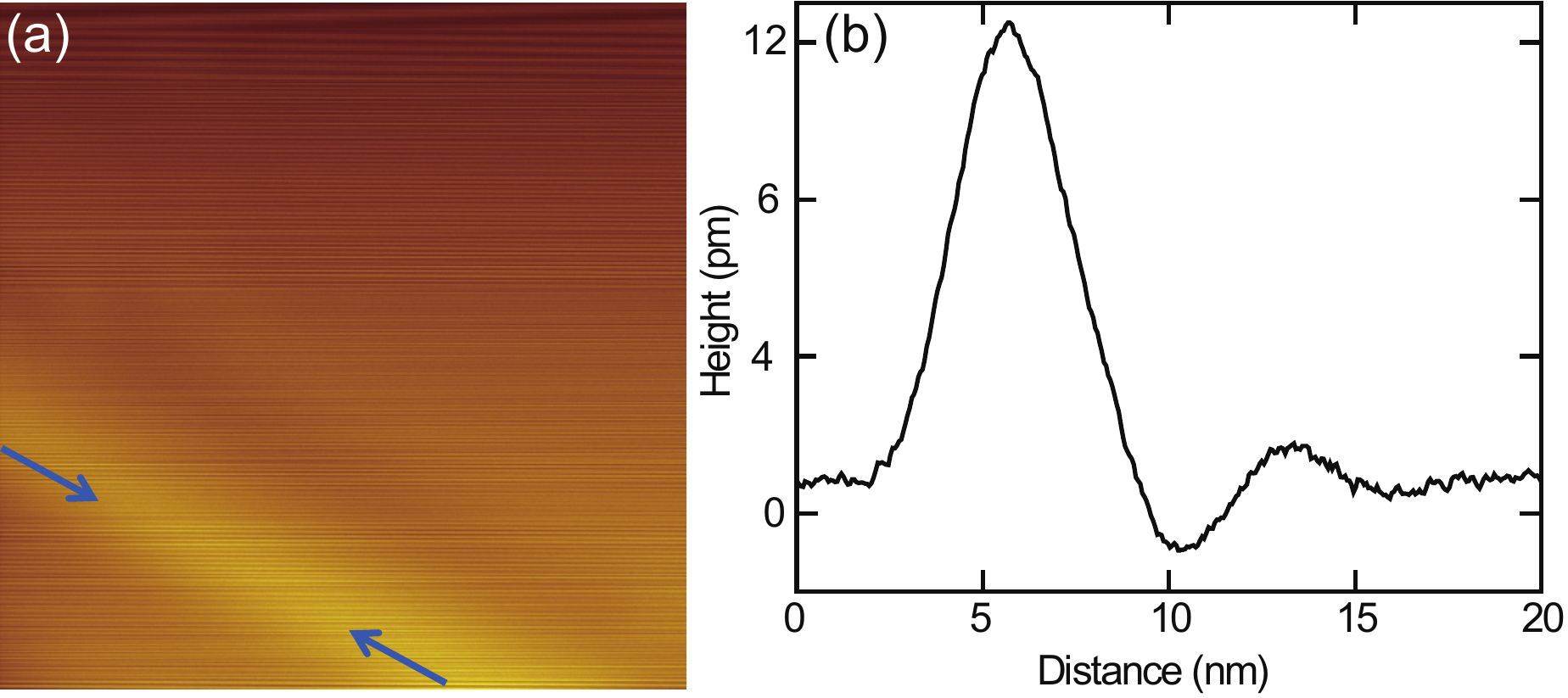}
\caption{(a) STM topography (25 nm $\times$ 25 nm) of a TB (blue arrows) at high voltage (\textit{V} = 1.0 V, \textit{I} = 48 pA). (b) Average linecut across the TB.}
\end{figure}

%